\documentclass[a4paper]{jpconf}
\usepackage{graphicx}
\begin{document}

\title{Numerical simulation of nonequilibrium states in a trapped 
Bose-Einstein condensate}

\author{A N Novikov$^{1,2}$, V I Yukalov$^{1,2}$ and V S Bagnato$^2$}

\address{$^1$ Bogolubov Laboratory of Theoretical Physics,
Joint Institute for Nuclear Research, Dubna 141980, Russia} 

\address{$^2$ Instituto de Fisica de S\~{a}o Carlos, Universidade de S\~{a}o Paulo, \\
CP 369, 13560-970 S\~{a}o Carlos, S\~{a}o Paulo, Brazil}

\begin{abstract}
In this work we present numerical study of a trapped Bose-Einstein condensate 
perturbed by an alternating potential. The relevant physical situation has been 
recently realized in experiment, where the trapped condensate of $^{87}$Rb, 
being strongly perturbed, exhibits the set of spatial structures. Firstly, 
regular vortices are detected. Further, increasing either the excitation 
amplitude or modulation time results in the transition to quantum vortex 
turbulence, followed by a granular state. Numerical simulation of the 
nonequilibrium Bose-condensed system is based on the solution of the  
time-dependent $3$D nonlinear Schr\"{o}dinger equation within the static and 
dynamical algorithms. The damped gradient step and time split-step Fourier 
transform methods are employed. We demonstrate that computer simulations 
qualitatively reproduce the experimental picture, and describe well the main 
experimental observables.
\end{abstract}

\section{Introduction}

As is well known, trapped Bose atoms in equilibrium at zero temperature form 
Bose-Einstein condensate (BEC) in the ground state. This system has been 
thoroughly studied theoretically as well as experimentally. For details we refer 
to the review articles and books \cite{Pit_Str,Peth_Sm,CBY,Yuk_basic} and 
references therein. If the condensate is slightly moved from equilibrium it 
usually exhibits collective excitations. Such dynamics is typical not only 
for BEC, but for the variety of quantum systems \cite{Bir_Naz,Yan_Lan,Saar_Reim}.

Nonequilibrium BEC can be created by injecting into the system energy, which
can be realized by applying an external modulation field. Two main ways of 
moving trapped BEC far from equilibrium are usually considered. The first one 
is a resonant excitation of coherent modes \cite{Yuk_non_gs, Yuk_coh_mod}. 
In this case the applied external perturbation is weak, but the modulation 
frequency of the alternating field is in resonance with one of the transition 
frequencies between topological modes. This method is a bit difficult for 
experimental realization, requiring fine tuning to resonance. The other way 
is to employ a strong external alternating field superimposed onto the 
stationary trap \cite{Bag_Yuk_progr}, where either the modulation time or 
excitation amplitude determine the energy injected into the system. In both 
these cases, BEC transfers from its ground to excited states, demonstrating 
a set of dynamical regimes.  The second approach has been recently realized 
in the experiment \cite{Vander_exp_1,Vander_exp_2}.

The experimental realization of nonequilibrium condensates stimulates interest 
to its numerical modeling and computer simulation \cite{Comp_meth}. Dealing 
with a trapped Bose-Einstein condensate, the nonlinear Schr\"oedinger equation 
(NLSE), which, in BEC physics sometimes is called the Gross-Pitaevskii 
equation \cite{Pit_Str}, is the main tool for the theoretical investigation. 
Nowadays, different numerical schemes \cite{Comp_meth,Taha} have been developed 
and successfully applied for the solution of second order partial differential 
equations, like NLSE. The methods differ in their stability, accuracy, 
computational cost, and ability of keeping the NLSE original properties. In 
general, all such methods could be classified as explicit, implicit and finite 
Fourier transform or pseudospectral methods. Each numerical scheme has its own 
merits and disadvantages, determining the accuracy of the NLSE solution. The 
presence of either random forces or potentials, being a prerequisit of 
nonequilibrium systems, complicates the numerical procedure, since the relevant 
NLSE solution is not smooth.

In our recent paper \cite{YNB} we presented the results of numerical simulation 
of a trapped condensate, under external perturbation, showing the appearance of
a granular state, analogous to that observed in the experiment. The simulations 
were based on the solution of the NLSE, with the parameters corresponding to 
the experimental setup. In the present paper we give the details of the 
accomplished numerical modeling. Also, we demonstrate that the chosen numerical 
methods qualitatively describe the typical physical picture, observed in the 
experiment.

The paper is organized as follows. In Sec. 1, we briefly review the experiment 
on the strong perturbation of a trapped BEC by external modulating field. 
Details of computer simulation are given in Sec. 2. In Sec. 3, we present the 
numerically simulated time-amplitude diagram.

\section{Experiment on perturbation of $^{87}$Rb} 

In the experiment \cite{Vander_exp_1,Vander_exp_2}, the $^{87}$Rb atoms (atomic 
mass $m = 1.444\time 10^{-22}$ g, scattering length $a_s = 0.557\times 10^{-6}$ cm), 
being cooled down to the temperature much lower than $T_c = 276$ nK, form the 
ground state Bose-Einstein condensate. The total number of atoms in the BEC 
fraction is $N\approx 1.5\times 10^5$. Stationary trapping potential has 
cylindrical symmetry, with the trap frequencies: $\omega_r = 2\pi\times 210$~Hz, 
$\omega_z = 2\pi\times 23$~Hz. After the stationary BEC state, obeying the 
Thomas-Fermi shape, is achieved, the trapping potential is modulated by an 
additional external alternating field, oscillating with the frequency 
$\Omega_0 = 2\pi\times200$~Hz. As a result, the total trap potential $V_{ext}$ 
can be approximated as \cite{Vander_exp_2}:

\begin{eqnarray}
\label{pot}
V_{ext}=\frac{1}{2}(\lambda^2[x\cos{\theta_1}+ y\sin{\theta_1}-z\sin{\theta_2}-\delta_1(1-\cos{\Omega_0t})]^2+ \nonumber\\
+[y\cos{\theta_1}-x\sin{\theta_1}-\delta_2(1-\cos{\Omega_0t})]^2+ \\
+[z\cos{\theta_2}+x\sin{\theta_2}-\delta_3(1-\cos{\Omega_0t})]^2), \nonumber
\end{eqnarray}
where $\lambda = \omega_r/\omega_z$, $\theta_i = A_i(1-\cos{\Omega_0t})$, 
$A_1 = \frac{\pi}{60}$, $A_2 = \frac{\pi}{120}$. The trapping frequency $\omega_r$ 
determines the oscillator length $l_r$ that defines the modulation parameters 
of the potential: $(\delta_1, \delta_2, \delta_3) = \alpha(2,5,3)$ $\mu$m /$l_r$, 
where $\alpha$ is an effective amplitude of pumping.  

If the external perturbation imposes a certain angular momentum onto the system, 
like in the case of the rotating \cite{rot_BEC} or laser-stirred condensate 
\cite{stir_BEC}, the formed vortices are aligned along the rotation axis. 
The number of vortices depends on the modulation frequency. The aligned vortices 
can form a completely ordered lattice. 

But in our case, as is seen from the expression (\ref{pot}), the modulation 
field used in the experiment does not impose a certain rotation axis. This 
feature makes the main difference of our experiment, as compared to the case
of rotating BEC. Contrary to the rotated BEC, the external perturbation, we use,
just shakes the condensate cloud, injecting the energy into the trap and thus 
generating the diverse coherent topological modes \cite{Bag_Yuk_progr}.

Mathematically, coherent topological modes are stationary solutions of NLSE, 
including the ground state and other excited states. Different modes correspond 
to rather different spatial distributions of the condensate density, forming a 
variety of spatial shapes. As a result, different structures can be generated
in a trapped condensate \cite{Yuk_non_gs,Yuk_coh_mod}.
 
First modes, generated in the experiment, are quantum vortices. Since the 
external field does not inject a rotation moment, the vortices appear as  
vortex-antivortex pairs with the winding numbers plus/minus one These modes 
are energetically the most stable. The number of created vortices depends on 
the injected energy, hence can be controlled by either the pumping amplitude 
or by the modulation time. At a fixed amplitude, almost a linear dependence 
of the vortex number from time is observed, until the distance between the
vortex centers is less than four coherence lengths, when the interaction 
energy of two vortices is much smaller than the vortex energy. When the number 
of vortices becomes sufficiently large, the transition to random vortex 
turbulence occurs. 

The continuous energy pumping into the BEC cloud creates more vortices than 
the trap can host, as a result of which the vortices start strongly interacting 
and colliding, which destroys the regime of vortex turbulence. Then the system 
passes to the next dynamical regime, where the condensate forms high-density 
grains surrounded by sufficiently rarified gas. It can be demonstrated \cite{YNB} 
that these grains are coherent objects.

\section{Numerical simulation}

Experimental characteristics are found to be in good agreement with analytical 
estimates \cite{Bag_Yuk_progr,YNB}. Now we aim at accomplishing numerical 
calculations for the same setup as in the experiment. In this section, we focus 
on the techniques of numerical modeling as applied to the solution of NLSE 
describing the condensate dynamics driven by external perturbations. The 
corresponding numerical simulation is realized in two steps: first, we calculate
the parameters of the initial conditions describing the stationary state of the 
BEC in the trap, and then we consider time propagation. The latter is performed 
via the solution of the time-dependent NLSE for the desired time interval. In 
both these cases, the full 3D geometry is considered, and the wave function is 
determined on the 3D space grid. Below, the main points of the accomplished 
numerical simulations are illustrated.

\subsection{Initial conditions}

First, the wave function, describing the condensate stationary state, has to 
be computed. To this end, we consider the eigenproblem:
\begin{equation}
\label{eig}
H(\varphi)\varphi_n=E_n\varphi_n,
\end{equation}
where the index $n$ enumerates quantum states and the corresponding wave 
functions $\varphi_n$ and energies $E_n$. In equation (\ref{eig}), the nonlinear 
Hamiltonian $H(\varphi)$ is typical for a trapped BEC:
\begin{equation}
\label{ham}
H(\varphi)=-\frac{\hbar^2}{2m}\nabla^2+V_{ext}+gN|\varphi|^2 \; ,
\end{equation}
where $g = \frac{4\pi\hbar^2a_{s}}{m}$ is the interaction strength depending 
on the scattering length $a_s$ and atomic mass $m$, $V_{ext}$ is the trapping 
potential. The parameters of the Hamiltonian (\ref{ham}) exactly correspond to 
the experimental setup. The BEC of $^{87}$Rb atoms with the related $a_s$ and 
$m$ are considered. The total number of condensed particles is $N = 1.5\times 10^5$. 
The external field $V_{ext}$ is represented by the total trap potential in form 
(\ref{pot}). Since we interested in stationary solutions, the external 
perturbation is switched off, so that $V_{ext} \equiv V_{ext}({\bf r}, t = 0)$.

The eigenproblem (\ref{eig}) can be solved by a variety of numerical schemes, 
but iterative methods are the most convenient. We used the damped gradient step 
method \cite{damp_1,damp_2}, where the initial guess $\phi_n$ is improved by 
the iteration steps. Mathematically, this can be represented in the form: 
\begin{equation}
\label{step_proc}
\phi_n^{(i+1)}=\mathcal{O}[\phi_n^{(i)}-
\mathcal{D}(H(\phi_n^i)-<H(\phi_n^i>)\phi_n^i] \; ,
\end{equation}
where $\mathcal{D}$ is a damping operator, the symbol $\mathcal{O}$ labels 
the orthonormalization of the set of functions $\phi_n$ at each iteration step 
marked as $i$. The Schmidt orthonormalization is usually employed. 
Hamiltonian (\ref{ham}) is recalculated after the each iteration step. The main 
problem of the numerical scheme (\ref{step_proc}) is the proper choice of the 
damping operator $D$ providing the convergence of the iteration procedure. The 
damped gradient step method operates with the damping operator in the form
\begin{equation}
\label{dam_op}
\mathcal{D}=\frac{\delta}{T+E_0} \; ,
\end{equation}
where $\delta$ is a parameter controlling the iteration step size, 
$T = -\frac{\hbar^2}{2m} \nabla^2$ is the operator of kinetic energy. Since
kinetic energy can happen to be close to zero, parameter $E_0$ stabilizes the 
iteration process.

If the procedure converges, the final set of $\phi_n$, obtained after a number 
of iteration steps, corresponds to the eigenfunctions $\varphi_n({\bf r})$ with 
the spectra $E_n$.

After the eigenproblem (\ref{eig}) is solved, the total wave function 
$\Psi({\bf r})$ can be written as a superposition of the states
\begin{equation}\label{tot_wf}
\Psi({\bf r})=\sum_nC_n\varphi_n({\bf r}) \; ,
\end{equation}
and then normalized to the total number of condensed atoms
\begin{equation}\label{norm}
N=\int{|\Psi({\bf r})|^2}d{\bf r} \; .
\end{equation}

At $t = 0$, there is no external perturbation, thus the condensate is in 
equilibrium state corresponding to the lowest energy $E_0$. To take this into 
account, we apply the condition
\begin{equation}
C_0~>>~C_{ n>0}
\end{equation}
to the notation (\ref{tot_wf}).

Representation (\ref{tot_wf}) of the wave function corresponds to its 
expansion over the basis of coherent topological modes \cite{Yuk_coh_mod}, 
which allows one to treat non-ground-state condensates \cite{Yuk_non_gs}.

\subsection{Time evolution}

The static solution is followed by the time propagation of the condensate wave 
function $\psi({\bf r})$. For this purpose, we solve numerically the nonlinear 
3D time-dependent NLSE
\begin{equation}
\label{GPE}
i\hbar\frac{\partial}{\partial t}\Psi({\bf r},t)=
[-\frac{\hbar^2}{2m}\nabla^2+V_{ext}({\bf r})+g|\Psi({\bf r},t)|^2]\Psi({\bf r},t) \; ,
\end{equation}
using the obtained $\Psi({\bf r})$ as the initial condition. Similarly to the 
previous case, the parameters of NLSE exactly correspond to the experimental 
setup. To solve NLSE, the time split-step Fourier transform (TSSFT) method 
\cite{Hardin,Devr} is chosen. The advantage of this numerical scheme 
\cite{Comp_meth} is that it is unconditionally stable and keeps the main 
properties of NLSE, such as mass conservation, time reversibility, i.e.,  
time-inverse invariance. Also, the method is second-order accurate in both 
time and space.

The main idea of the TSSFT is in splitting the time-propagation between the 
kinetic and potential components of the total energy. Both these components 
can be approximated by exponents related to the solutions of the linear 
Schr\"odinger equation (see Ref. \cite{Devr} for details). Mathematically, 
the mechanism of time propagation has the following form:
\begin{equation}
\label{TSSFT}
\Psi({\bf r},t+\Delta t)\approx\exp{(i T\Delta t/2)}
\exp{(-i V \Delta t)}\exp{(i T\Delta t/2)}\Psi({\bf r},t) \; ,
\end{equation}
where $T = -\frac{\hbar^2}{2m}\nabla^2$ and 
$\widehat V = V_{ext} + g|\Psi({\bf r},t)|^2$ are the operators of kinetic 
and potential energy, respectively.

The relation between the exponential operators obeys the Baker-Cambell-Hausdorf 
theorem, which states that $e^Ae^B = e^C$ is valid if $C = A + B + [A,B] +...$. 
Since the TSSFT method skips the commutation relations of the kinetic and 
potential operators, numerical error grows with time. The symmetrical form of 
(\ref{TSSFT}), with respect to the operators, allows to remove the first 
commutator term, increasing the accuracy of time propagation \cite{Fleck}. 

The exponential derivative operator is easily calculated in the momentum space 
rather than in the coordinate space, because of the well known relation
\begin{equation}
\label{ft}
\exp{\left(\frac{d^2}{d{\bf q}^2}\right)}\Psi({\bf q},t)=
\exp{(-k^2)}\mathcal{F}[\Psi({\bf q},t)] \; ,
\end{equation}
where symbol $\mathcal{F}$ means the Fourier transform, $k$ is a variable 
conjugated to $q$.

Taking into account expressions (\ref{TSSFT}) and (\ref{ft}), the full time 
propagation of the condensate wave function for the time step $\Delta t$ is 
performed in three stages:
\begin{enumerate}
\item
The Fourier transform of the initial function $\Psi({\bf r},t)$ is followed by 
the half time-step $\Delta t/2$, with the operator of kinetic energy in the 
momentum space. The first stage is completed by the inverse Fourier transform 
that results in the new function $\Psi({\bf r},t)\to\tilde\Psi({\bf r},t)$.
\item
Recalculation of the potential operator $V$ with the function 
$\tilde\Psi({\bf r},t)$ and further full-time step propagation of 
$\tilde\Psi({\bf r},t)$, with the recalculated term $V$. As a result, the 
transformation $\tilde\Psi({\bf r},t)\to\hat\Psi({\bf r},t)$ is obtained.
\item
The final stage equals the first step, but with the function  
$\hat\Psi({\bf r},t)$. This completes the time propagation, since 
$\hat\Psi({\bf r},t)\to\Psi({\bf r},t + \Delta t)$ \; . 
\end{enumerate}

The time step $\Delta t$ should be small enough to keep the accuracy of time 
propagation. The increment $\Delta t$ fulfills the condition \cite{Feit}
\begin{equation}
<\Psi|H|\Psi>~<<~\frac{2\pi}{\Delta t} \; .
\end{equation}

\subsection{Other numerical aspects}

Dealing with a grid-based technique, the parameters of the computation box 
are extremely important. The size of the computational box strictly depends 
on the trapping frequencies, total number of condensed particles and the kind 
of boundary conditions. The proper size of the 3D grid avoids the unphysical 
truncation of the operated wave function and its reflection from the grid 
boundaries during the perturbation. The later is important, if reflecting 
boundary conditions are employed. In the present simulation, we used the 
reflecting boundary conditions for both, the static and dynamical simulations. 
In the experiment, the applied external perturbation heats up the BEC leading 
to the transition from condensed to thermal phase for a part of atoms. This 
process is almost regular in time. Consequently, the number of condensed 
particles in the trap permanently decreases. To follow the experiment, we 
introduced into the system a small dissipation replacing the term 
$i\hbar\frac{\partial}{\partial t}\Psi({\bf r},t)$ in the NLSE (\ref{GPE}) 
by $(i-\gamma)\hbar\frac{\partial}{\partial t}\Psi({\bf r},t)$, where $\gamma$ 
is a phenomenological term. The same approach is also used by the other 
authors \cite{Tsubota}. In simulations, during the $60$ ms of perturbation, 
approximately 15 percents of the initial condensed particles are lost, which
is in good agreement with the experiment. Of course, introducing the dissipation 
term $\gamma$ does not conserve the initial norm of the wave function 
(\ref{norm}), which implies the diminishing number of condensed atoms.

\section{Results and discussion}

Computer simulations reproduce well the experimental results. The applied 
external perturbation, modelled by (\ref{pot}) creates, first, dipole 
oscillations of the condensate cloud, followed by vortex structures 
corresponding to vortex-antivortex pairs. The vortex nature of the observed 
structures is confirmed by quantized circulation having the values plus or 
minus one. The circulation is calculated by the method proposed in 
Ref \cite{Foster}. Increasing either the amplitude of pumping or the 
excitation time produces more and more vortices, creating the random vortex 
tangle, representing, according to Feynman \cite{Feynman}, quantum turbulence. 
The next regime observed in the simulations is grain turbulence, where 
continuous trap modulation destroys all the vortices, forming high-density 
grains surrounded by rarified gas. The typical physical picture, obtained in
numerical simulations, is represented by the time-amplitude diagram shown 
in Fig. 1. The experimental diagram, published in Ref \cite{Vander_exp_2}, 
is in qualitative agreement with that obtained in the numerical simulation. 
The area inside the rectangle reproduces the experiment in the best way.

\begin{figure}[ht]
\centering
\includegraphics[width=12cm]{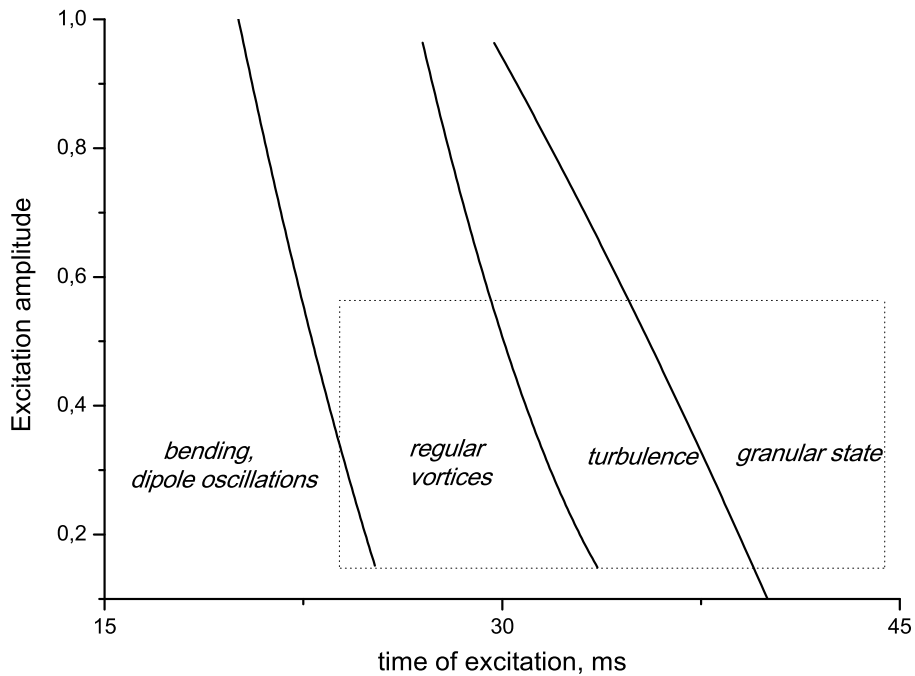}
\caption{The time-amplitude diagram, as found in numerical simulations. The 
area inside the rectangle is in the best agreement with the experiment.}
\end{figure}

Numerical simulations give very good quantitative values for the main 
characteristics observed in the experiment and also found from theoretical
arguments. Thus, the critical number of vortices $N_{cr}\approx 25$, required
for creating the regime of quantum turbulence, is the same in the experiment, 
numerical simulations, as well as in theoretical estimates. The grains can 
be considered as coherent objects, since their typical linear size practically
coincides with the coherence length, as is found in the experiment and in the 
simulation.  

In numerical simulations, we also observe one more regime \cite{YNB}, arising 
after the granular state, the regime of wave turbulence, when all BEC is 
destroyed, and turbulence is formed by small-amplitude waves. Here we compare 
the experimental and numerical results, because of which the regime of wave 
turbulence is not included in the diagram, since it has not yet been reached 
in experiment.

\section{Acknowledgments}

Financial support from FAPESP and CNPq (Brazil) is appreciated. Two of the 
authors (V.I.Y. and A.N.N.) acknowledge financial support from the Russian 
Foundation for Basic Research, under project 14-02-00723.

\end{document}